\documentclass[twocolumn,prl,aps]{revtex4-2}
\usepackage{floatflt}
\usepackage{psfrag}
\usepackage{amsmath}   
\usepackage{amssymb}
\usepackage{amsfonts}
\usepackage{mathrsfs}
\usepackage{graphicx}

\usepackage[dvipsnames]{xcolor}
\usepackage[colorlinks=true, citecolor=Cerulean,linkcolor=blue, urlcolor = Cerulean]{hyperref}

\begin{document}

%\title{Discovery of Slot Plasma Excitations in a AlGaN/GaN Plasmonic Crystal}

%\author{A.~R.~Khisameeva$^{a}$, A.~Shuvaev$^{a}$, P.~A.~Gusikhin$^{b}$, A.~S.~Astrakhantseva$^{c}$, A.~Pimenov$^{a}$,  I.~V.~Kukushkin$^{c}$, V.~M.~Muravev$^{c}$ \footnote{Corresponding E-mail: muravev@issp.ac.ru}}
%\affiliation{$^a$ Institute of Solid State Physics, Vienna University of Technology, 1040 Vienna, Austria \\
%$^c$ Institute of Solid State Physics, RAS, Chernogolovka, 142432 Russia}
\title{Discovery of Slot Plasma Excitations in a AlGaN/GaN Plasmonic Crystal}

\author{A.~R.~Khisameeva$^{a}$, A.~Shuvaev$^{a}$, I.M.~Moiseenko$^{b}$, P.~A.~Gusikhin$^{c}$, A.~S.~Astrakhantseva$^{c}$, A.~Pimenov$^{a}$,  D.A.~Svintsov$^{b}$, I.~V.~Kukushkin$^{c}$, V.~M.~Muravev$^{c}$ \footnote{Corresponding E-mail: muravev@issp.ac.ru}}
\affiliation{
$^a$ Institute of Solid State Physics, Vienna University of Technology, 1040 Vienna, Austria \\
$^b$ Laboratory of 2d Materials for Optoelectronics, Moscow Institute of Physics and Technology, Dolgoprudny, 141700 Russia\\ 
$^c$ Institute of Solid State Physics, RAS, Chernogolovka, 142432 Russia}

\date{\today}

\begin{abstract}
We experimentally investigate the terahertz spectrum of plasma excitations in a plasmonic crystal based on AlGaN/GaN two-dimensional electron system (2DES). While screened plasmon modes with linear dispersion are readily observed in the plasmonic crystals, the existence of unscreened modes localized in the slots between the gates has remained unobserved until now. We discover this slot plasma excitation exhibiting square-root dispersion. It turned out that these slot plasmons follow an unconventional wave-vector quantization rule,  $q_u=(N + 1/4) \times \pi/l_u$ for even integers $N$, and require the condition for excitation $q_u h \ll 1$, where $h$ is the gate-to-2DES distance and $l_u$ is the slot width. We develop an analytical model that accurately captures the found dispersion and relaxation, revealing a non-trivial $-\pi/4$ phase shift upon plasmon reflection at the gate edge. Experiments demonstrate that the slot plasmons persist up to room temperature, thereby enabling a broad range of opportunities for the advancement of plasmonic devices.
\end{abstract}

\maketitle

A plasmonic crystal consists of a two-dimensional electron system (2DES) with a periodic gate structure placed on the surface above the 2DES. Groundbreaking studies on plasma waves within semiconductor-based two-dimensional systems were first performed with plasmonic crystals over four decades ago~\cite{Allen:1977, Theis:1977, Heitmann:1982}. From that point onward, this configuration has served as a key foundation for uncovering a number of novel physical phenomena, such as the detection of minigaps within the plasmon dispersion\cite{Heitmann:1984, Heitmann:1986}, two-dimensional plasmons in stratified materials\cite{Wang:2011, Basov:2018}, Tamm states in plasmonic crystals\cite{Shaner:2012, Shaner:2013}, and diverse phases of plasmonic crystals\cite{Shur:2010, Knap:2023, Knap:2024}. In addition, the majority of emerging plasmonic devices are founded on the plasmonic crystal concept, encompassing detectors for terahertz (THz) radiation\cite{Allen:2002, Knap:2009, Lusakowski:2014}, oscillators and amplifiers operating in the THz range\cite{Tsui:1980, Mikhailov:1998, Popov:2020}, and THz phase shifters\cite{Muravev:2022, Muravev:2023}.

An incident electromagnetic wave excites standing plasma waves in the plasmonic crystal~\cite{Allen:1977, Theis:1977, Heitmann:1982}. The currently prevailing theoretical model suggests that dispersion of the plasma waves is analogous to the Kronig-Penney model, which describes the particle behavior in a periodic potential~\cite{Aizin:2012, Popov:2015, Kachorovskii:2012, Svintsov:2019, Kachorovskii:2024}. It turned out that the model agrees well with experimental data in the broad range of plasmonic crystal parameters~\cite{Svintsov:2019, Khisameeva:2025}. For the interested reader we refer to the Supplemental Material for details of this theoretical model~\cite{Supplemental}. When the distance from the gate to the 2DES is relatively small, the tight-binding Kronig–Penney model predicts two distinct plasma modes in the plasmonic crystal. The first is an ordinary screened plasmon, which is excited in the region directly underneath the gate, with its spectrum taking the linear form~\cite{Chaplik:1972}
\begin{equation}
\omega_g(q) = \sqrt{\frac{n_s e^2 h}{m^{\ast} \varepsilon \varepsilon_0}} \, q \qquad (q h \ll 1),
\label{scr_plasmon}
\end{equation} 
where $n_s$ is the 2D electron density, $m^{\ast}$ is the effective electron mass, $h$ is the distance between the gate and the 2DES, $l_g$ is the width of the gate strips, the wave vector
$q_g=M \times \pi/l_g$ ($M=1,2,\ldots$ is an integer), and $\varepsilon$ is the dielectric permittivity of the semiconductor crystal, specifically $\varepsilon_{\rm GaN} = 12.8$. This gated plasmon mode has been observed in a number of experimental studies~\cite{Shur:2010, Knap:2023, Knap:2024, Khisameeva:2025}.

Importantly, the theory also predicts a second class of standing plasma waves in the plasmonic crystal. This mode corresponds to an ordinary, unscreened plasmon excited in the gap between the gates, with its frequency described by~\cite{Stern:1967}
\begin{equation}
    \omega_u(q) = \sqrt{\frac{n_s e^2}{2 m^{\ast} \varepsilon(q) \varepsilon_0} \, q}.
\label{plasmon}
\end{equation} 
Here, $q_u = N \times \pi/l_u$ ($N=1,2,\ldots$ is an integer number, $l_u$ is the width of the slot between the gates) is the plasmon wave vector, $\varepsilon(q)$ is the effective dielectric permittivity of the medium surrounding the 2DES. Unfortunately, this plasmon mode has never been observed in experiments with plasmonic crystals.

In the present work, we investigate the terahertz spectrum and relaxation of plasma excitations in a plasmonic crystal based on an AlGaN/GaN heterostructure. Our experimental data reveal that the plasmonic crystal supports two distinct series of resonant plasmon modes. The first corresponds to screened plasmons with linear dispersion (\ref{scr_plasmon}), localized in the semiconductor under the gate. Crucially, we have discovered a previously unobserved branch of unscreened plasmon modes localized in the inter-gate slots. These “slot” plasmons exhibit a square-root dispersion (\ref{plasmon}) and obey an unconventional quantization condition:
\begin{equation}
q_u=\left (N+ \frac{1}{4} \right) \times \frac{\pi}{l_u} \qquad  N=2,4,6 \ldots 
\end{equation} 
which originates from a nontrivial phase shift of $-\pi/4$ acquired upon plasmon reflection from the gate edge. We develop an analytical model that quantitatively captures both the dispersion and the relaxation of these slot modes, and identify their key excitation criterion 
$q_u h \ll 1$. Our results pave the way for engineering of new THz plasmonic devices.

\begin{figure}[t!]
    \centering
    \includegraphics[width=0.8\linewidth]{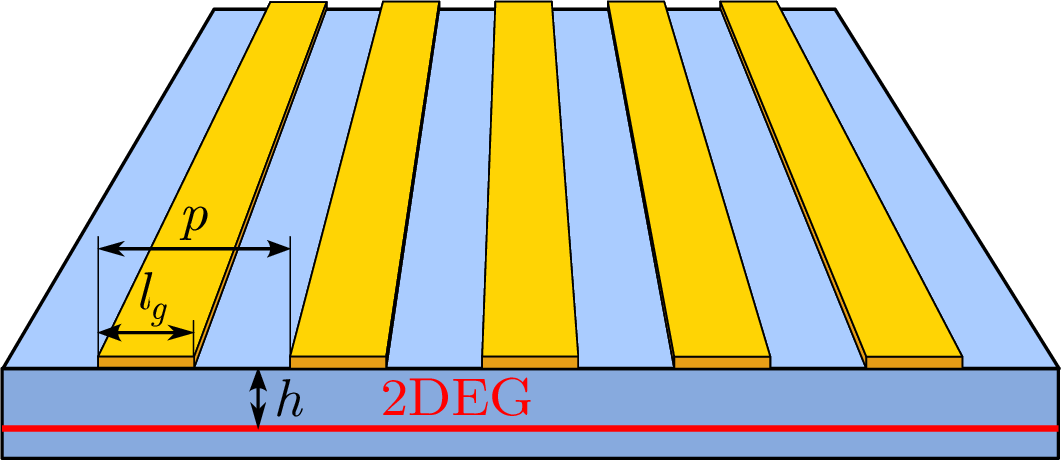}
    \caption{A schematic of the plasmonic crystal.}
    \label{1}
\end{figure}

In our study, we have explored the samples which were fabricated from the Al$_{0.22}$Ga$_{0.78}$N/GaN heterostructure grown by the molecular beam epitaxy on a SiC substrate. The two-dimensional electron layer was separated from the surface by the $7$~nm SiC, $2$~nm GaN, and $25$~nm Al$_{0.22}$Ga$_{0.78}$N cap layers.  In total, $h=34$~nm. The density of two-dimensional electrons in the heterojunction was $n_{s}=9.6\times 10^{12}~\text{cm}^{-2}$ with  mobility $\mu = 7 500~\text{cm}^2/\text{V$\cdot$s}$, derived from low-temperature ($T=5$~K) transport experiments. The same parameters for the ambient ($T=300$~K) conditions were  $n_{s}=1.2\times 10^{13}~\text{cm}^{-2}$ with  mobility $\mu = 1 950~\text{cm}^2/\text{V$\cdot$s}$. A Cr($24$~nm)/Au($280$~nm) grid-shaped gate is evaporated on the top surface of the sample (Fig.~\ref{1}). We fabricated samples with grating periods varying from $p=8$~\textmu{}m up to $24$~\textmu{}mand a filling factor (the ratio of stripe width to period) ranging from $f=0.35$ to $0.83$. A detailed description of all sample parameters and heterostructure composition is provided in the Supplemental Material~\cite{Supplemental}. The experiments were conducted using the time-domain spectroscopy setup, which covers a frequency range from $0.15$ to $4$~THz. The spectrometer is equipped with a continuous-flow liquid-helium cryostat with a superconducting solenoid ($\pm 7$~T) and transparent windows. All measurements are performed at the base sample temperature of $T=5$~K. To minimize noise from water absorption lines in the air, the THz beam path is enclosed in an evacuated chamber. Spectroscopy measurements are carried out using a $6$~mm aperture positioned close to the sample surface to ensure that the electromagnetic radiation passes only through the grating-gate active region of the plasmonic crystal. The polarization of the electromagnetic radiation is directed perpendicular to the gate grids.

\begin{figure}[t!]
    \centering
    \includegraphics[width=0.9\linewidth]{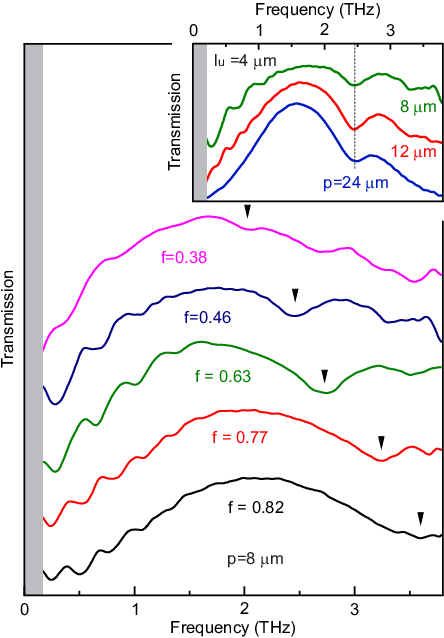}
    \caption{Transmission spectra of the plasmonic crystal for various grid filling factors 
    $f=0.38, 0.46, 0.70, 0.77$ and $0.82$, corresponding to slot widths $l_u = 5.0, 4.3, 2.4, 1.8$, and $1.4$~\textmu{}m, respectively. The crystal period is fixed at $p = 8$~\textmu{}m. The arrows point to the frequency positions of the $N=2$ slot plasmon mode. The inset displays transmission spectra for three samples with identical slot widths $l_u = 4.0$~\textmu{}m, but different periods: $p = 8$~\textmu{}m (green curve), $12$~\textmu{}m (red curve), and $24$~\textmu{}m (purple curve).}
    \label{2}
\end{figure}

Figure~\ref{2} shows transmission spectra measured on a set of samples with different slot widths while the lattice period was held fixed at $p = 8$~\textmu{}m. For clarity the traces have been offset vertically. Each spectrum contains two distinct series of plasmon resonances. The evenly spaced, lower‑frequency series is identified as screened plasmon modes with wave vector $q_g=M \times \pi/(l_g +2h)$ ($M=1,3, 5 \ldots$ is an odd integer).
Here, we find that the incident electromagnetic wave selectively excites only optically active (bright) plasmon modes, as confirmed by the dispersion analysis presented below. The most striking feature in Fig.~\ref{2} is the emergence of an additional series of resonances in the high-frequency spectral range. The fundamental plasmon resonance of this series (indicated by the arrow) shifts to lower frequency as the slot width increases, which implies that the associated plasmon mode is localized in the ungated gap between gate fingers. To verify this hypothesis we compared three samples with the same slot width $l_u = 4.0$~\textmu{}m, but different sizes of the gated regions of the plasmonic crystal: $l_g = 4$~\textmu{}m (blue curve),  $l_g = 8$~\textmu{}m (red curve), and $20$~\textmu{}m (purple curve). Increasing $l_g$ leaves the resonance frequency unchanged, confirming that this mode is localized in the slot rather than in the gated region.

In order to better understand the discovered slot plasmon modes, we develop a two-stage analytical theory based on the solution of Maxwell's equations in the gated 2DES with an isolated slot.
Details can be found in the accompanying paper~\cite{Svintsov:2026} and Supplemental Material~\cite{Supplemental}. At the first stage, we establish the phase of plasmon reflection at an individual gate edge using the Wiener-Hopf approach to the diffraction problem~\cite{Kay1959, Rejaei2015}. The phase appears equal to $-\pi/4$ in the experimentally relevant limit $q_uh\ll 1$, while the absolute value of the reflection coefficient is order of unity. At the second stage, we introduce the complex reflection coefficient into the Fabry-Perot-like dispersion relation for the two-edged slot $r^2\exp\{2iq_ul_u\}=1$. As a result, the wave vectors of cavity eigenmodes is quantized according to the following unconventional rule
\begin{equation}
q_u=\left (N+ \frac{1}{4} \right) \times \frac{\pi}{l_u} \qquad  N=2,4,6 \ldots ,
\label{quantization}
\end{equation}
while the relation between frequency $\omega$ and wave vector $q_u$ is that of ordinary ungated plasmon (\ref{plasmon}).

\begin{figure}[!t]
    \centering
    \includegraphics[width=\linewidth]{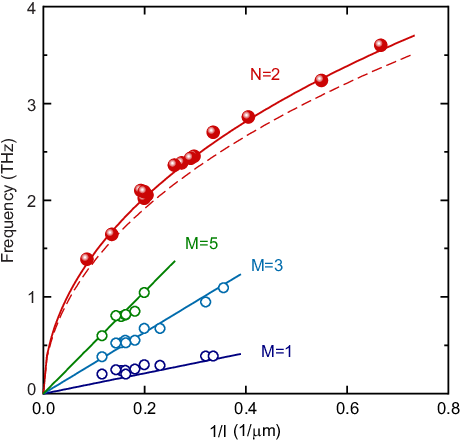}
    \caption{Frequency of the fundamental slot‑plasmon mode ($N=2$) versus the inverse slot width, $1/l_u$. Solid red dots: experimental resonance frequencies. Solid red line: theory from Eq.~(\ref{plasmon}) using the wave vector $q_u=(2 \pi + \pi/4)/l_u$. Dashed red line: Eq.~ (\ref{plasmon}) with standard quantization rule $q_u = 2 \pi/l_u$. Empty symbols: the frequencies of the screened plasmon modes with $M=1$ (empty purple dots), $M=3$ (empty blue dots), and $M=5$ (empty green dots). Straight lines: theoretical prediction from Eq.~(\ref{scr_plasmon}) with $q_g=M \times \pi/l_g$.}
    \label{3}
\end{figure}

To validate the proposed theoretical model, Fig.~\ref{3} shows the measured frequency of the fundamental slot‑plasmon mode ($N=2$) plotted against the inverse slot width, $1/l_u$. 
The solid red curve is the prediction of Eq.~(\ref{plasmon}) using the wave-vector quantization condition $q_u=(2 \pi + \pi/4)/l_u$, including the non‑trivial phase shift of $-\pi/4$. In these calculations we used an electron effective mass $m^{\ast} = 0.24 \, m_0$, determined independently from measurements in a magnetic field~\cite{Supplemental}. The effective dielectric constant, $\varepsilon (q)$, was calculated taking into account the layered AlGaN/GaN heterostructure architecture~\cite{Volkov:1988}. There is a remarkable agreement between the experimental data and theory that takes into account a non-trivial $-\pi/4$ phase shift. For comparison, the dashed red curve in the same panel shows the result obtained with the conventional quantization rule $q_u = 2 \pi/l_u$. The discrepancy between this curve and the data highlights the importance of the additional phase shift. Measurements performed in a perpendicular magnetic field reveal hybridization between the slot‑plasmon mode and the cyclotron resonance. The cyclotron frequency is given by $\omega_c = e B/m^{\ast}$, and the hybridized mode follows the relation $\omega^2 = \omega_u^2 + \omega_c^2$, where $\omega_u$ is the zero‑field plasmon frequency. Representative transmission spectra in the presence of the magnetic field and the extracted magnetodispersion are provided in the Supplemental Material~\cite{Supplemental}.

To clarify the physical origin of the low‑frequency resonance series shown in Fig.~\ref{2}, we plot the measured frequencies of the $M = 1$, $M=3$, and $M = 5$ resonances versus $1/l_g$ (empty symbols in Fig.~\ref{3}). The experimental points fall on straight lines when plotted against $1/l_g$, consistent with the acoustic-like dispersion expected for the screened plasmons~\cite{Chaplik:1972}. The solid purple, blue and green lines in Fig.~\ref{3} are theoretical plots of Eq.~(\ref{scr_plasmon}) using $q_g=M \times \pi/l_g$. There is good agreement between experimental data and theory. 

\begin{figure}[!t]
    \centering
    \includegraphics[width=0.9\linewidth]{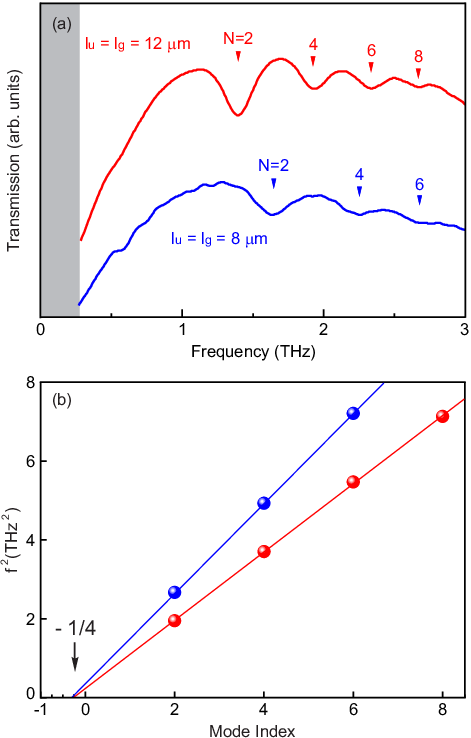}
    \caption{(a) Representative transmission spectra for crystals with equal ungated and gated widths $l_u=l_g=12$~\textmu{}m (red) and $l_u=l_g=8$~\textmu{}m (blue), showing the fundamental slot mode and its higher‑frequency harmonics. (b) Squared resonance frequency $f_N^2$ plotted versus mode index $N$. The linear dependence confirms the unscreened plasmon dispersion [Eq.~(\ref{plasmon})]. Experimentally observed bright resonances correspond to even mode indices $N = 2, 4, 6, 8$. The data reveal a nontrivial phase offset of 1/4 in the mode sequence.}
    \label{4}
\end{figure}

Transmission spectra in Fig.\ref{2} for large slot widths, $l_u$, reveal clear harmonics of the slot-plasma excitations. To explore this in detail, we analyze representative spectra from plasmonic crystals with $l_u=l_g=12$~\textmu{}m (red curve in Fig.~\ref{4}(a)) and $l_u=l_g=8$~\textmu{}m (blue curve in Fig.~\ref{4}(a)). The fundamental slot mode and its higher-frequency harmonics are well resolved in the transmission data. To assign mode indices $N$ to these resonances, Fig.~\ref{4}(b) plots the squared resonance frequency $f_N^2$ versus the mode index $N$. The points follow a straight line in accordance with unscreened plasmon spectrum (\ref{plasmon}), and the bright resonances observed experimentally correspond to even modes, $N=2, 4, 6, 8$. The plot also indicates a nontrivial phase offset of $-\pi/4$ in the mode sequence.

Another matter of interest is associated with the relaxation dynamics (damping) of the slot plasmons excited in a plasmonic crystal. For conventional 2D plasmons the quality factor is set primarily by carrier scattering in the 2DES and can be estimated as $Q=\omega_p \tau$, where $\omega_p$ is the plasmon angular frequency and $\tau$ is the carrier momentum-relaxation time. Using the transport mobility of the AlGaN/GaN heterostructure studied here yields $\tau$ that corresponds to $Q \approx 20$ at the plasmon frequency $\omega_p/2 \pi = 3$~THz. This estimate substantially overestimates the experimentally observed $Q$ of the slot-plasmon resonance, which is $Q \approx 5$ at the same frequency (Fig.~\ref{2}). The discrepancy is corroborated by the much sharper screened-plasmon resonances in Fig.~\ref{2} compared with the broad slot-plasmon feature. This observation reveals a remarkable property of the slot-plasmon modes: they are quasi‑bound and leak energy into the plasmons beneath the gates. Physically this appears as a finite decay rate even for an ultraclean 2DES~\cite{Svintsov:2026}, with
\begin{equation}
Q \sim \frac{1}{\sqrt{q_u h}}.    
\label{Q}
\end{equation}
Because $Q$ decreases only slowly with increasing $h$, the decay rate remains significant unless the gate–2DES separation, $h$, is made extremely small. Thus observing slot modes requires small $q_u h$ factor, which in turn demands an extremely small gate–2DES separation. Presumably this factor prevented the observation of slot plasmons in the recent detailed study of a plasmonic crystal fabricated from an AlGaAs/GaAs heterostructure~\cite{Khisameeva:2025}, since the gate–to–2DES separation, $h$, in those devices is substantially larger than in the AlGaN/GaN structures considered here.

\begin{figure}[!t]
    \centering
    \includegraphics[width=\linewidth]{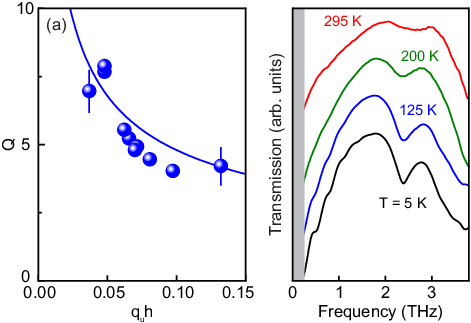}
    \caption{(a) Dependence of the fundamental slot‑plasmon quality factor $Q$ on the dimensionless parameter $q_u h$. Experimental data (points) and the best fit of $Q \sim 1/\sqrt{q_u h}$ (solid blue curve) are shown. $Q$ increases as the wave vector $q_u$ decreases. (b) Temperature dependence of the transmission spectra for the plasmonic crystal with $l_u=4$~\textmu{}m and $l_g=6$~\textmu{}m, measured at $T = 5, 125, 200$ and $295$~K.}
    \label{5}
\end{figure}

Figure~\ref{5}(a) shows how the fundamental slot-plasmon quality factor $Q$ depends on the parameter $q_u h$. As the wave vector $q_u$ decreases, the slot-plasmon $Q$ increases. The solid black curve in Fig.~\ref{5}(a) is the best fit of Eq.~(\ref{Q}) to the experimental data. For the AlGaN/GaN device under study the fitted relation is $Q = 1.5/\sqrt{q_u h}$. Closely related to relaxation is the temperature dependence of the plasmonic response. Figure~\ref{5}(b) presents transmission spectra measured at $T=5, 125, 200$, and $295$~K for the plasmonic crystal with $l_u=4$~\textmu{}m and $l_g=6$~\textmu{}m. The slot plasmon remains essentially unchanged up to $T=125$~K, then its amplitude decreases gradually. It is important to note that plasmon resonance is still observable under ambient conditions. Notably, the plasmon linewidth is nearly constant up to room temperature, which indicates that slot-plasmon relaxation is dominated by decay into the gated regions of the plasmonic crystal.

It is important to distinguish two fundamentally different states of the plasmonic crystal: weak-binding and tight-binding regimes. The regime is governed by the dimensionless parameter $Z$~\cite{Aizin:2012, Svintsov:2019, Kachorovskii:2024}
\begin{equation}
Z= \frac{1}{2} \left( \frac{q_g}{q_u} + \frac{q_u}{q_g} \right) \approx \frac{1}{2} \left( \sqrt{q_u h} + \frac{1}{\sqrt{q_u h}} \right).
\end{equation}
Weak binding is realized for $q_u h \sim 1$ (parameter $Z \sim 1$), whereas tight binding corresponds to $q_u h \ll 1$, when $Z \gg 1$ is large. Consequently, the slot-plasmon mode can be excited only in the tight-binding regime of the plasmonic crystal. It is absent in the weak-binding regime, such as that realized in a plasmonic crystal fabricated from an AlGaAs/GaAs heterostructure~\cite{Khisameeva:2025}.

In conclusion, we have experimentally discovered and theoretically explained a previously unobserved family of slot (unscreened) plasmon modes in an AlGaN/GaN plasmonic crystal.
It turned out that the discovered slot modes have a number of outstanding physical properties.
They exhibit the square-root dispersion of unscreened two-dimensional plasmons and obey an unconventional quantization rule $q_u=(N + 1/4) \times \pi/l_u$ with even $N$, rather than the integer quantization typical of Fabry–Perot optical cavities. Observation of these modes requires the tight-binding condition $q_u h \ll 1$, which is satisfied in our AlGaN/GaN devices but was not in earlier AlGaAs/GaAs experiments with much larger gate–to–2DES separation. Developed analytical model quantitatively reproduces the measured dispersion and shows that slot plasmons acquire a reflection phase of $- \pi/4$. This phase shift is the physical origin of the $1/4$ offset in the quantization rule and is essential for correct mode indexing and frequency prediction. Crucially, we demonstrate that these modes persist up to room temperature, opening a pathway to robust and compact terahertz plasmonic components that operate under ambient conditions.

\begin{acknowledgments}
    This work was supported as a part of the ISSP RAS State assignment. 
\end{acknowledgments}

\bibliography{main}

\end{document}